\RequirePackage{ifpdf}
\ifpdf 
\documentclass[pdftex]{sigma}
\else
\documentclass{sigma}
\fi

\numberwithin{equation}{section}

\begin{document}

\allowdisplaybreaks

\renewcommand{\PaperNumber}{068}

\FirstPageHeading

\renewcommand{\thefootnote}{$\star$}

\ShortArticleName{Hidden Symmetries of Stochastic Models}

\ArticleName{Hidden Symmetries of Stochastic Models\footnote{This paper is a
contribution to the Proceedings of the O'Raifeartaigh Symposium on
Non-Perturbative and Symmetry Methods in Field Theory (June
22--24, 2006, Budapest, Hungary). The full collection is available
at \href{http://www.emis.de/journals/SIGMA/LOR2006.html}{http://www.emis.de/journals/SIGMA/LOR2006.html}}}

\Author{Boyka ANEVA}

\AuthorNameForHeading{B. Aneva}

\Address{Institute for Nuclear Research and Nuclear Energy,
Bulgarian Academy of Sciences,\\
 72 Tsarigradsko chaussee, 1784 Sof\/ia, Bulgaria}
 \Email{\href{mailto:blan@inrne.bas.bg}{blan@inrne.bas.bg}}

\ArticleDates{Received November 23, 2006, in f\/inal form May
04, 2007; Published online May 18, 2007}

\Abstract{In the matrix product states approach to $n$ species
dif\/fusion processes the statio\-nary probability distribution is
expressed as a matrix product state with respect to a~quadratic
algebra determined by the dynamics of the process. The quadratic
algebra def\/ines a noncommutative space with a $SU_q(n)$ quantum
group action as its symmetry. Boundary processes amount to the
appearance of parameter dependent linear terms in the algebraic
relations and lead to a reduction of the $SU_q(n)$ symmetry. We
argue that the boundary operators of the asymmetric simple
exclusion process generate a tridiagonal algeb\-ra whose irriducible
representations are expressed in terms of the Askey--Wilson
polyno\-mials. The Askey--Wilson algebra arises as a symmetry of the
boundary problem and allows to solve the model exactly.}


\Classification{60J60; 17B80}

\section{Introduction}

 Stochastic reaction-dif\/fusion processes are of both
theoretical and experimental interest not only because
they describe various mechanisms in physics and chemistry
\cite {schu1} but they also provide a way of modelling
phenomena like traf\/f\/ic f\/low \cite {schre}, kinetics
of biopolimerization \cite {mac}, interface growth
\cite {spo}.

  A Markov process can be described
in terms of a master equation for the
probability
distribution $P(s_i, t)$ of a
stochastic variable
$s_i = 0, 1, 2,\dots,n-1$
at a site $i = 1,2,\dots,L$ of a linear chain. A state
on the lattice at a time $t$ is determined by
the occupation numbers $s_i$ and a transition
to another conf\/iguration $s_i'$ during an
inf\/initesimal time step $dt$ is given by the
pro\-ba\-bility $\Gamma(s, s')dt$. The rates
$\Gamma \equiv \Gamma ^{ik}_{jl}$, $i,j,k,l =0, 1,2,\dots,n-1$
are assumed to be independent from the position in
the bulk. At the boundaries, i.e.\ sites $1$ and $L$
additional processes can take place with rates
$L_i^j$ and $R_i^j$. Due to probability conservation
\begin{gather*}
 \Gamma (s,s) =
   -\sum_{s'\neq s} \Gamma (s',s).
\end{gather*}
The master equation for the
time evolution of a stochastic system
\begin{gather*}
 \frac {dP(s,t)}{dt} = \sum _{s'} \Gamma (s,s')P(s',t)
\end{gather*}
is mapped to a
Schr\"odinger equation for a
quantum Hamiltonian in imaginary time
\begin{gather*}
  \frac {dP(t)}{dt}=-HP(t),
\end{gather*}
where
\begin{gather*}
  H= \sum_j H_{j,j+1}+ H^{(L)} + H^{(R)}.
\end{gather*}
The probability distribution
thus becomes a state
vector in the conf\/iguration space
of the quantum spin chain and the ground state of the
Hamiltonian, in general non-Hermitian,
corresponds to the steady state of the stochastic dynamics
where all probabilities are stationary.
The mapping provides a connection with integrable
quantum spin chains and allows for exact results of
the stochastic dynamics with the formalism of quantum
mechanics.

  Examples are the dif\/fusion processes of
particles hopping with hard core repulsion between
lattice sites
$i$, $j$ with probability rates $g_{ij}$ and subjected to ref\/lecting
(or free) boundary conditions \cite {schu2}.
In the symmetric case $g_{ij}=g_{ji}$ the stochastic Hamiltonian is
the $SU(2)$ symmetric spin $1/2$
isotropic Heisenberg ferromagnet
\begin{gather*}
   H=-\frac {1}{2}\sum_i (\sigma^x_i \sigma^x_j+
  \sigma^y_i\sigma^y_j+\sigma ^z_i\sigma^z_j -1).
\end{gather*}
The $SU(2)$ symmetry, yet unrevealed in the original
master equation becomes manifest through the mapping
and allows for exact results of the stochastic dynamics.
The asymmetric exclusion process is the driven dif\/fusive lattice gas of
particles hopping with rates
$\frac {g_{i,i+1}}{g_{i+1,i}} = q \neq 1$
and is mapped to a $SU_q(2)$-symmetric $XXZ$ chain
with anisotropy $\Delta = \frac {(q+q^{-1})}{2}$. 

\section[Matrix product state approach to diffusion models]{Matrix product state approach to dif\/fusion models}

An algebraic approach, developed for the study of the steady state
properties of Markov processes and related to the quantum Hamiltonian
picture is the matrix-product-state ansatz \cite{der1, der2}.
The idea is that the stationary probability
distribution, i.e.\ the ground state of the quantum Hamiltonian
with nearest-neighbour interaction in the bulk and single-site
boundary terms is expressed
as a product of (or a trace over) matrices
that form a representation of a quadratic algebra
\[
   \Gamma ^{ik}_{jl}D_iD_k=x_lD_j-x_jD_l,
 \qquad i,j,k=0,1,\dots,n-1,
\]
determined by the dynamics of the process.

  In the following we consider dif\/fusion processes with $n$
species on a chain of $L$ sites with nearest-neighbour
interaction with exclusion, i.e.\ a site can be
either empty or occupied by a~particle of a given type.
In the set of occupation numbers $(s_1,s_2,\dots,s_L)$
specifying a conf\/igu\-ration of the system $s_i=0$ if
a site $i$ is empty, $s_i=1$ if there is a f\/irst-type
particle at a site $i,\dots,s_i=n-1$ if there is
an $(n-1)$th-type particle at a site $i$.
On successive sites the species $i$ and $k$
exchange places with probability $g_{ik}dt$, where
$i,k=0,1,2,\dots,n-1$. With $i<k$, $g_{ik}$
are the probability rates of hopping to the left,
and $g_{ki}$
to the right. The event
of exchange happens if out of two adjacent sites one is
a vacancy and the other is occupied by a particle, or
each of the sites is occupied by a particle of a
dif\/ferent type.
The $n$-species symmetric simple exclusion process is
known as the lattice gas model of particle hopping
between nearest-neighbour sites with a constant rate
$g_{ik}=g_{ki}=g$. The $n$-species asymmetric simple
exclusion process with hopping in a preferred
direction is the driven dif\/fusive lattice gas of
particles moving under the action of an external f\/ield.
The process is totally asymmetric if all jumps occur in
one direction only, and partially asymmetric if there is
a dif\/ferent non-zero probability of both left and
right hopping. The number of particles ${n_i}$ of
each species in
the bulk is conserved
$\sum\limits_{i=0}^{n-1} n_i = L
$
and this is the case of
periodic boundary conditions. In the case of open
systems, the lattice gas is coupled to external reservoirs
of particles of f\/ixed density. In most studied examples
\cite{der2, do}
one considers phase transitions inducing \cite {kru} boundary
processes when a particle of type $k$,
$k=1,2,\dots,n-1$ is added with
a rate $L_k^0$ and/or removed with a rate $L_0^k$ at
the left end of the chain, and it is removed with a rate
$R_0^k$ and/or added with a rate $R_k^0$ at the right
end of the chain.

   For dif\/fusion processes the transition rate matrix becomes
simply
 $\Gamma ^{ik}_{ki}=g_{ik}$ and the $n$-species dif\/fusion algebra
\cite {r} has the form
\begin{gather}
  g_{ik}D_iD_k - g_{ki}D_kD_i = x_kD_i - x_iD_k,\label{6}
\end{gather}
where $g_{ik}$ and $g_{ki}$ are positive
(or zero) probability rates,
 $i,k=0,1,\dots,n-1$ and $x_i$ satisfy
\[
 \sum_{i=0}^{n-1} x_i=0.
\]
(No summation over repeated indices in equation~\eqref{6}.)

The quadratic algebra has a Fock representation in an
auxiliary Hilbert space where the $n$ generators act as
operators. For systems with periodic boundary conditions,
the stationary probability distribution is related to the
expression
\begin{gather}
  P(s_1,\dots,s_L)={\rm Tr}\,(D_{s_1}D_{s_2}\cdots D_{s_L}).\label{7}
\end{gather}
When
boundary processes are considered the stationary probability
distribution is related to a~matrix element in the auxiliary vector space
\begin{gather}
 P(s_1,\dots,s_L)=\langle w\vert D_{s_1}D_{s_2}\cdots D_{s_L}\vert v\rangle \label{8}
\end{gather}
with respect to the vectors $\vert v\rangle $ and $\langle w\vert $,
def\/ined by
 the boundary conditions
\begin{gather*}
  \langle w\vert (L_i^kD_k + x_i)=0,\qquad
   (R_i^kD_k - x_i)\vert v\rangle=0,
\end{gather*}
where the $x$ sum up to zero, because of the
form of the boundary rate matrices
\begin{gather*}
  L_i^i=-\sum_{j=0}^{L-1} L^i_j,\qquad
  R_i^i=-\sum_{j=0}^{L-1} R^i_j.
\end{gather*}
These relations simply mean that
one associates with an
occupation number $s_i$ at position $i$ a~matrix
$D_{s_i}=D_k$ ($i=1, 2,\dots,L$;
 $k=0, 1,\dots,n-1$) if
a site $i$ is occupied by a $k$-type particle.
The number of all possible conf\/igurations of an
$n$-species stochastic system on a chain of $L$
sites is $n^L$ and this is the dimension
in the conf\/iguration space of the stationary
probability distribution as a state vector. Each
component of this vector, i.e.\ the (unnormalized)
steady-state weight of a given conf\/iguration, is
a trace or an expectation value in the
auxiliary space given by \eqref{7} or~\eqref{8}. The
quadratic algebra reduces the number of
independent components to only monomials
symmetrized upon using the relations~\eqref{6}.

The algebra admits an involution,
hence the matrices $D_i$
 can be chosen Hermitian
\begin{gather*}
  D_i=D_i^+, \qquad g_{ik}^+=g_{ki}, \qquad
  x_i=-x_i^+
\end{gather*}
(or $D_i=-D_i^+$, if $g_{ik}=g_{ki}^+$).

  The relations \eqref{6} allow an ordering of the elements
$D_k$. Monomials of given order are the Poincar\'e--Birkhof\/f--Witt (PBW) basis in the algebra. The quadratic algebra
generated
by the $n$ elements $D_k$ obeying the $n(n-1)/2$
relations \eqref{6} is an associative algebra with a unit $e$.
The ordered monomials
\begin{gather*}
   D_{s_1}^{n_1}D_{s_2}^{n_2}\cdots D_{s_l}^{n_l},
\end{gather*}
where $s_1<s_2<\cdots <s_l$, $l \geq 1$ and
$n_1, n_2,\dots, n_l$ are non-negative integers, are a
linear basis in the algebra.

 Thus, to f\/ind the stationary probability
distribution
one has to compute traces or matrix elements
with
respect to the vectors $\vert v\rangle$ and $\langle w\vert $
of  monomials of the form
  $D_{s_1}^{n_1}D_{s_2}^{n_2}\cdots D_{s_L}^{n_L}$.
The problem to be solved is twofold -- to f\/ind a
representation of the matrices $D$ that is a
solution of the quadratic algebra
and match the
algebraic solution with the boundary conditions.
Finding a~representation of the quadratic algebra and
solving the corresponding boundary problem allows one to
conclude about the symmetries of the model.

  In the case of symmetric dif\/fusion
with all $x_i$ distinct the
dif\/fusion algebra is of Lie-algebra type.
The $n$ generators $D_i$, and $e$ can be
mapped \cite {an} to the generators $J_{jk}$ of
$SU(n)\times U(1)$ and the mapping is
invertible. The universal enveloping algebra generated
by $D_i$ belongs to the UEA
of the Lie-algebra of $SU(n)\times U(1)$.

  In the known example of exactly solvable $2$- and
$3$-species models \cite{der2, ess, ar} of asymmetric
dif\/fusion through the matrix product ansatz,
the solution of the quadratic algebra is
provided by a deformed bosonic oscillator algebra, if
both $g_{ik}$ and $g_{ki}$ dif\/fer from zero,
or by inf\/inite-dimensional matrices, if $g_{ik}=0$.
In the general $n$ case, because of the ordering
procedure, the solution of the quadratic algebra has
to be consistent with the diamond lemma in ring theory,
also known as the
braid associativity condition in quantum groups.
As shown in \cite{r, an} if all
parameters $x_i$ on the RHS of equation~\eqref{6} are
equal to zero, the homogeneous
quadratic algebra def\/ines a multiparameter quantized
non-commutative space  realized
equivalently as a $q$-deformed Heisenberg
algebra \cite{za, we} of $n$ oscillators
depending on $n(n-1)/2 + 1$ parameters
(in general on $n(n-1)/2+n$ parameters):
\begin{gather*}
  a_ia_i^+-r_ia_i^+a_i=1 ,\qquad
  a_i^+a_j^+-q_{ji}a_j^+a_i^+=0 , \\
  a_ia_j - q_{ji}a_ja_i=0, \qquad
  a_ia_j^+-q_{ji}^{-1}a_j^+a_i=0,
\end{gather*} where $i<j$; $i, j = 0, 1,\dots,n-1$, the deformation
parameters $r_i, q_{ij}$ are model-dependent parameters given in
terms of the probability rates. The associative algebra generated by
the elements $D_i$ in this case  belongs to the universal enveloping
algebra of the multiparameter deformed Heisenberg algebra to which a
consistent multiparameter $SU_q(n)$ quantization corresponds. Thus
the bulk symmetry of a process without boundary conditions is
$SU_q(n)$ and the process is equivalent to an integrable spin chain.
For a non-homogeneous algebra with $x$-terms on the RHS of~\eqref{6}, only
then is braid associativity satisf\/ied if, out of the coef\/f\/icients
$x_i$, $x_k$, $x_l$ corresponding to an ordered triple $D_iD_kD_l$, either
one coef\/f\/icient $x$ is zero or two coef\/f\/icients $x$ are zero, and
the rates are respectively related. The dif\/fusion algebras in this
case can be obtained by either a change of basis in the
$n$-dimensional non-commutative space or by a~suitable change of
basis of the lower-dimensional quantum space realized equivalently
as a~lower-dimensional deformed Heisenberg algebra. The appearance
of the non-zero linear terms in the RHS of the quantum plane
relations due to boundary processes leads to a lower-dimensional
non-commutative space and a reduction of the $SU_q(n)$ invariance.

\section{The open asymmetric exclusion process}

   We consider now the two-species partially
asymmetric simple exclusion process with incoming
and outgoing particles at both boundaries. We simplify the
notations, namely, at the left bounda\-ry
a particle can be added with probability $\alpha dt$
and removed with probability $\gamma dt$, and at the
right boundary it can be removed with probability
$\beta dt$ and added with probability $\delta dt$.
The system is described by the conf\/iguration
set $s_1,s_2,\dots,s_L$ where $s_i=0$ if a site
$i=1,2,\dots,L$ is empty and $s_i=1$ if a site $i$
is occupied by a particle. The particles hop
with a probability $g_{01}dt$ to the left and
with a probability $g_{10}dt$ to the right,
where without
loss of generality we can choose the right
probability
rate $g_{10}=1$ and the left probability rate
$g_{01}=q$. The model depends on f\/ive parameters -- the bulk
probability rate $q$ and the four boundary rates. The totally
asymmetric process corresponds to $q=0$.
The quadratic algebra
\begin{gather*}
D_1D_0-qD_0D_1=x_1D_0-D_1x_0, \qquad x_0+x_1=0
\end{gather*}
is solved \cite {ev} by a pair of deformed oscillators $a$, $a^+$
(and $x_0=-x_1=1$)
\begin{gather*}
  D_0=\frac {1}{1-q} + \frac {a^+}{\sqrt {1-q}},  \qquad
  D_1=\frac {1}{1-q} + \frac {a}{\sqrt {1-q}}.
\end{gather*}
The boundary conditions have the form
\begin{gather}
  (\beta D_1-\delta D_0)\vert v\rangle =\vert v\rangle,\qquad
  \langle w\vert (\alpha D_0 - \gamma D_1)=\langle w\vert.\label{16}
\end{gather}
For a
given conf\/iguration $(s_1, s_2,\dots,s_L)$ the
stationary probability is given by
the expectation value
\begin{gather*}
  P(s)=\frac {\langle w\vert D_{s_1}D_{s_2}\cdots D_{s_L}
   \vert v\rangle}{Z_L},
\end{gather*}
where
$D_{s_i}=D_1$ if a site $i=1,2,\dots,L$ is occupied and
$D_{s_i}=D_0$ if a site $i$ is empty and
$Z_L=\langle w \vert (D_0+D_1)^L \vert v\rangle$
is the normalization factor to the stationary probability
distribution.
Once the representation of the dif\/fusion algebra and the boundary
vectors are known, one can evaluate all the physical quantities such
as the current $J$ through a bond between site $i$ and site $i+1$,
which has a very simple form 
\begin{gather*}
  J=\frac {Z_{L-1}}{Z_L},
\end{gather*}
the mean density  at a site $i$
\begin{gather*}
  \langle s_i\rangle =\frac {\langle
   w\vert (D_0+D_1)^{i-1}D_1(D_0+D_1)^{L-i}
   \vert v\rangle}{Z_L},
\end{gather*}
the two-point correlation function
\begin{gather*}
  \langle s_is_j\rangle =\frac {\langle w\vert
  (D_0+D_1)^{i-1}D_1(D_0+D_1)^{j-i-1}D_1
  (D_0+D_1)^{L-j}\vert v\rangle}{Z_L}
\end{gather*}
and higher correlation functions.

   The matrix-product algebraic relations provide solvable
(and representation independent) recursions for the
stationary probability, the current and the correlation
functions. These recursions have been obtained in earlier
works \cite{do, san}, however they were not readily generalized to
other models. The algebraic matrix-product
method allows for generalizations to the multispecies
case \cite{der2} and was also  to the dynamical process~\cite{sti}.

  To solve the boundary problem for a process with
only incoming particles at the left boundary and only
outgoing particles at the right one ($\delta = \gamma =0$
in \eqref{16})
we choose the vector
$\vert v\rangle$ to be the (unnormalized!)
eigenvector of the annihilation
operator $a$ for a real value of the parameter $v$ and
the vector $\langle w\vert$ to be the eigenvector
(unnormalized and dif\/ferent from the conjugated one)
of the creation
operator for the real parameter $w$:
\begin{gather}
  \vert v\rangle
 =e_q^{-\frac {1}{2}vw}e_q^{va^+}\vert 0\rangle,
  \qquad \langle w\vert
  =\langle 0 \vert e_q^{wa}e_q^{-\frac {1}{2}wv},\label{21}
\end{gather}
where
$e_q^z=\sum \frac {z^n}{[n]!}$ and
$[n]=\frac {1-q^n}{1-q}$. The factor
$e_q^{-\frac {1}{2}vw}$ in \eqref{21} is due to the
condition $\langle w\vert v\rangle =1$, which is a
convenient choice in physical applications.
According to the algebraic solution, these are also
eigenvectors of the shifted operators $D_0$, $D_1$ with the
corresponding relations of the eigenvalues
\begin{gather*}
 \frac {1}{\alpha}=\frac {1}{1-q}+\frac {w}{\sqrt {1-q}}, \qquad
 \frac {1}{\beta} =\frac {1}{1-q}+\frac {v}{\sqrt {1-q}}.\label{22}
\end{gather*}
The choice of the boundary vectors to be
deformed coherent states \cite{aco, ku}  of
the $q$-deformed oscillator \cite{ku, chai} set
underlying the algebraic solution simplif\/ies
the procedure of obtaining all the relevant physical quantities. One
has to merely normally order all the expressions -- monomials in $a$, $a^+$
and use the coherent states properties. The  representation
has the proper $q=0$ limit which allows to obtain all the physical quantities
of the totally asymmetric process as well. The deformed coherent state
solution in the
case of only incoming particle to the left and only outgoing particle
to the right provides the most simple and convenient approach to a~unif\/ied
description of both the partially and the totally asymmetric processes.

  The asymmetric exclusion process was studied and analyzed in relation to
orthogonal polynomials technique. In the case of only injected
particles at the left boundary and only removed at the right one the
exact solution was related to the $q$-Hermite \cite{ev} and
Al-Salam--Chihara polynomials \cite{sa}. In the general case of injected
and removed particles at both boundaries with four nonzero boundary
parameters the exact solution was shown to be related to the
Askey--Wilson polynomials \cite{wa}. Our analysis aims to further
stress the relation to the orthogonal polynomials. We put an
emphasis on the important characteristic of the open asymmetric
exclusion process that the bulk properties in the stationary state
strongly depend on the boundary parameters. We f\/ind that the
boundary operators generate the Askey--Wilson tridiagonal algebra
whose irreducible modules are the Askey--Wilson polynomials. The
boundary tridiagonal algebra reveals algebraic properties of the
asymmetric open exclusion process which are deeply  related to the
Askey--Wilson polynomials and allow for the exact solution of the
model.

\section{Symmetry of the open asymmetric exclusion process}

  In the general case of incoming and outgoing particles at both boundaries
there are four opera\-tors $\beta D_1$, $-\delta D_0$, $-\gamma D_1$,
$\alpha D_0$ and one needs a pair of deformed oscillators to form two
linear independent boundary operators acting on the dual vector
spaces. To solve this problem within the matrix product ansatz we
consider a particular form of a $q$-oscillator algebra, known in the
literature as $CU_q(2)$ (or $(u,u)$, $u>0$) algebra. It is generated
by three elements with the def\/ining commutation relations
\begin{gather*}
 [N,
A_{\pm}]= \pm A_{\pm}, \qquad [A_-, A_+]=uq^N+uq^{-N} 
\end{gather*}
and a central element 
\begin{gather*}
   Q=A_+A_- + u\frac {q^N -q^{1-N}}{1-q}.
\end{gather*} This is an associative algebra and any invertible
transformation of the generators is admissible. It can be presented
in equivalent forms by means of the transformations 
\begin{gather*}
  \hat A_+ =q^{N/2}A_+ , \qquad \hat A_- =A_-q^{N/2}
\end{gather*}
with
\begin{gather*}
 [\hat A_-, \hat A_+]_{q^{-1}}= uq^{2N+1/2}+uq^{1/2}
\end{gather*}
or
\begin{gather*}
  \tilde A_+ =q^{-N/2}A_+ , \qquad \tilde A_- =A_-q^{-N/2}
\end{gather*}
with
\begin{gather*}
 [\tilde A_-, \tilde A_+]_q= uq^{-1/2} +uq^{-2N-1/2}.
\end{gather*}
We consider now the two sets of operators
\begin{gather*}
 a_1=q^{N/2}A_+ + q^N ,\qquad
 a^+_1=q^{-N/2}A_+ + \frac {1}{1-q}q^{-N}
\end{gather*}
and
\begin{gather*}
 a_2  = A_-q^{N/2}+\frac {1}{1-q^{-1}}q^{N},\qquad
 a^+_2 = A_-q^{-N/2} + q^{-N}
\end{gather*}
obeying
\begin{gather*}
  a_1a^+_1 -qa^+_1a_1=1, \qquad  a_2a^+_2- q^{-1}a^+_2a_2=1.
\end{gather*}
This representation allows one to write the boundary operators
in the equivalent form
\begin{gather}
\beta D_1 -\delta D_0   \nonumber\\  
\qquad{}=\frac {- x_1\beta}{1-q}q^{N/2}A_+ +
 x_0\delta q^{-1}A_-q^{N/2} + \left(\frac{-x_1\beta}{(1-q)^2}
+ x_0\delta q^{-1}\right)q^N - \frac {x_1\beta}{1-q} -
 \frac {x_0\delta q^{-1}}{1-q^{-1}},\nonumber     \\ 
\alpha D_0 - \gamma D_1 \label{35}\\
\qquad{}=
 x_0\alpha q^{-N/2}A_+ -
\frac {x_1\gamma q^{-1}}{1-q^{-1}}A_-q^{-N/2}+
\left(x_0 \alpha - \frac {x_1\gamma q^{-1}}{(1-q^{-1})^2}\right)q^{-N}
+\frac {x_0\alpha}{1-q} + \frac {x_1\gamma q^{-1}}{1-q^{-1}}.\nonumber
\end{gather}
We separate the shift parts from the boundary operators. Denoting
the corresponding rest operator parts by $A$ and $A^*$ we write the
left and right boundary operators in the form
\begin{gather*}
\beta D_1 -\delta D_0 = A -\frac {x_1\beta}{1-q} -
 \frac {x_0\delta q^{-1}}{1-q^{-1}},\qquad
\alpha D_0 - \gamma D_1 = A^* +\frac {x_0\alpha}{1-q} +
 \frac {x_1\gamma q^{-1}}{1-q^{-1}}.
\end{gather*}

\begin{proposition} The operators $A$ and $A^*$ defined by
\begin{gather*}
 A=  \beta D_1 -\delta D_0 + \frac {x_1\beta}{1-q}
 +\frac {x_0 \delta q^{-1}}{1-q^{-1}},\qquad
 A^*= \alpha D_0 - \gamma D_1- \frac {x_0\alpha}{1-q}
   -\frac {x_1\gamma q^{-1}}{1-q^{-1}}
\end{gather*}
and their $q$-commutator
\begin{gather*}
 [A,A^*]_q=q^{1/2}AA^*-q^{-1/2}A^*A
\end{gather*}
form a closed linear algebra
\begin{gather}
[[A,A^*]_q,A]_q = -\rho A^* -\omega A - \eta,\qquad 
[A^*,[A,A^*]_q]_q = -\rho^*A -\omega A^* -\eta^*,
\label{39}
\end{gather}
where the structure constants are given by
\begin{gather}
 -\rho=\beta \delta q^{-1}x_1^2(q^{1/2}+q^{-1/2})^2, \qquad
-\rho^*=\alpha \gamma q^{-1}x_0^2(q^{1/2}+q^{-1/2})^2,
\label{40}
\\
 -\omega= (-x_0x_1)q^{-1}\left((\beta +
\delta(q^{1/2}-q^{-1/2})^2\right)
\left(\alpha (q^{1/2}-q^{-1/2})^2 +\gamma\right) \nonumber\\ 
\phantom{-\omega=}{}+(-x_0x_1)q^{-1}(q^{1/2}-q^{-1/2})^2QK ,  \label{41}\\  
 \frac {-\eta}{q-q^{-1}}=  x_1\beta x_0\delta q^{-1}Q\left(
  x_0\alpha +\frac {-x_1\gamma q^{-1}}{(1-q^{-1})^2}\right)
 +K\left( \frac {-x_1\beta}{(1-q)^2}+x_0\delta q^{-1}\right) ,\nonumber \\  
 \frac {-\eta^*}{q-q^{-1}}= K\left( x_0\alpha + \frac {-x_1\gamma q^{-1}}
 {(1-q^{-1})^2}\right) -x_0\alpha x_1\gamma q^{-1}Q
 \left( \frac {-x_1\beta}{(1-q)^2} +x_0\delta q^{-1}\right),\nonumber
\end{gather}
where
\begin{gather}
 K=(-x_ox_1)q^{-1}\left( \alpha \delta -
\frac {\beta \gamma}{(q^{1/2}-q^{-1/2})^2}\right). \label{42} 
\end{gather}
\end{proposition}

The proposition is straightforward to verify by using the
representation for $A$ and $A^*$ on the RHS of formula~\eqref{35}.
Relations \eqref{39} are the well known  Askey--Wilson relations 
\begin{gather*}
 A^2A^*-(q+q^{-1})AA^*A +A^*A^2 = \rho A^* + \omega A  +\eta,\nonumber \\ 
 A^{*2}A -(q+q^{-1})A^*AA^*+AA^{*2}= \rho^* A +\omega A^* +\eta^*
\end{gather*}
for the shifted boundary operators $A$, $A^*$. The algebra~\eqref{39} was f\/irst
considered in the works of Zhedanov \cite{zhe1, zhe2} and
recently discussed in
a more general framework of a tridiagonal algebra~\cite{ter1, ter2}.
This is an associative algebra
with a unit generated by a (tridiagonal) pair of operators $A$, $A^*$
and def\/ining relations
\begin{gather}
 [A, A^2A^*- \beta' AA^*A +A^*A^2 - \gamma' (AA^*+A^*A)-\rho A^*]= 0,\nonumber  \\  
 [A^*, A^{*2}A - \beta' A^*AA^*+AA^{*2}- \gamma^{'*} (AA^*+A^*A)-\rho^* A]= 0.
\label{44}
\end{gather}
In the general case a tridiagonal pair is determined by the sequence of
scalars $\beta'$, $\gamma'$, $\gamma^{'*}$, $\rho$, $\rho^*$ from a f\/ield $K$.
Note that we keep the conventional notations $\beta$, $\gamma$, $\rho$
for the scalars def\/ining a~tridiagonal pair in equation~\eqref{44}, however
we use them with the index $'$ to avoid confusion with the
boundary rates.
Tridiagonal pairs have been classif\/ied according to the dependence on
the scalars \cite{ter1}. Examples are the $q$-Serre relations with
 $\beta' =q+q^{-1}$ and
$\gamma' = \gamma^{'*} = \rho = \rho^* =0$
\begin{gather*}
  [A, A^2A^*- (q+q^{-1})AA^*A +A^*A^2]= 0,\qquad 
 [A^*, A^{*2}A -(q+q^{-1})A^*AA^*+AA^{*2}]= 0
\end{gather*}
and the Dolan--Grady relations \cite {dol} with
$\beta'=2$, $\gamma' = \gamma'^{*} =0$, $\rho = k^2$, $\rho^* =k^{*2}$
\begin{gather*}
[A,[A,[A,A^*]]]=k^2[A,A^*],\qquad
[A^*,[A^*,[A^*,A]]]=k^{*2}[A^*,A].
\end{gather*}
Tridiagonal pairs are determined up to an af\/f\/ine transformation
\begin{gather*}
 A \rightarrow tA+c, \qquad A^* \rightarrow t^*A^*+c^*,
\end{gather*} 
where $t$, $t^*$, $c$, $c^*$ are some scalars. The af\/f\/ine
transformation can be used to bring a tridiagonal pair in a
canonical form with $\gamma' = \gamma'^{*} =0$.

The boundary operators of the asymmetric exclusion process
obeying the Askey--Wilson algebra \eqref{39} form a tridiagonal pair with
$\beta' = q+q^{-1}$, $\gamma' = \gamma'^{*} =0$, and $\rho$, $\rho^*$,
$\omega = \omega^*$, $\eta$, $\eta^*$ given by equations \eqref{40}--\eqref{42}.
The Askey--Wilson algebra possesses some important properties that allow to
obtain its ladder representations, spectra, overlap functions.
This fact immediately
provides a solution to the boundary problem. We brief\/ly touch the spectral
problem
following the argumentation of Zhedanov et al.\ (for details
see \cite{zhe1, zhe2}).
Let $f_r$ be an eigenvector of $A$ with
eigenvalue $\lambda_r$
\begin{gather*}
 Af_r=\lambda _r f_r.
\end{gather*}
Then we can construct a new eigenstate
\begin{gather*}
 f_s=(A g(A)+A^*h(A)+A_0k(A))f_r,
\end{gather*} 
where $A_0$ denotes the $q$-commutator $[A,A^*]_q$,
and \begin{gather*}
 Af_s=\lambda_sf_s.
\end{gather*}
It follows from the algebra that $f_s$ will also be an eigenvector of $A$, if
for the new eigenvalue the quadratic relation holds
\begin{gather*}
 \lambda_r^2 +\lambda_s^2 -(q+q^{-1})\lambda_r \lambda_s -\rho=0.
\end{gather*}
This yields for each state $f_r$ two neighbouring states whose eigenvalues are
the roots of the above quadratic equation.
In this parametrization the operator $A$ is diagonal and the operator $A^*$ is
tridiagonal
\begin{gather*}
 Af_r=a_{r+1}f_{r+1}+b_rf_r+c_{r-1}f_{r-1}.
\end{gather*}
The expressions for the spectrum and the matrix coef\/f\/icient can be obtained
explicitly, e.g.\ the spectrum reads (with $x_1^2 =1$)
\begin{gather*}
\lambda_r = q^{-1/2}(\beta \delta)^{1/2}
\frac {q^r+q^{-r}}{q^{1/2}- q^{-1/2}}.
\end{gather*}
The right boundary vector $\vert v \rangle$ can be thus related to a suitably
chosen eigenstate of the
shifted operator $A$.
The algebra possesses a duality property. Due to the duality property
the dual basis exists in which
the operator $A^*$ is diagonal and the operator $A$ is tridiagonal.
The left boundary vector $\langle w \vert$ will be related to an
eigenstate of the shifted operator $A^*$ in the dual basis. We have
\begin{gather*}
 A^*f_p^*=\lambda^*f^*_p,\qquad
 Af^*_s=a^*_{s+1}f^*_{s+1}+b^*f^*_s +c^*_{s-1}f^*_{s-1}.
\end{gather*}
The overlap function of the  two basis
$\langle s \vert r \rangle = \langle f^*_s \vert f_r \rangle$
is expressed in terms of the Askey--Wilson polynomials.
To make the relation more transparent we  rescale the generators
\begin{gather*}
 A \rightarrow \big(q^{1/2}-q^{-1/2}\big)
\frac {1}{\beta}\sqrt {\frac {\gamma}{\alpha}} A, \qquad
A^* \rightarrow \big(q^{1/2}-q^{-1/2}\big)
\frac {q^{1/2}}{\sqrt {\alpha \gamma}}A^*.
\end{gather*}
The tridiagonal relations for the transformed operators read
\begin{gather*}
[A, A^2A^* -(q +q^{-1})AA^*A+A^*A^2 +abcdq^{-1}(q-q^{-1})^2A^*]=0,\nonumber  \\   
[A^*,A^{*2}A -(q+q^{-1})A^*AA^* +AA^{*2}+(q-q^{-1})^2A]=0,
\end{gather*}
where $abcd=\frac {\gamma}{\alpha} \frac {\delta}{\beta}$.
Let $p_n=p_n(x;a,b,c,d)$ denote the $n$th Askey--Wilson
polynomial \cite{ask}
depending on four parameters $a$, $b$, $c$, $d$
\begin{gather}
p_n(x;a,b,c,d) = {}_4\Phi_3 \left ( \begin {array}{c}
   q^{-n},abcdq^{n-1},ay,ay^{-1} \\
     ab,ac,ad \end{array} \vert\, q;q \right )
 \label{57}
\end{gather}
with $p_0=1$, $x=y+y^{-1}$ and $0<q<1$. On the RHS of equation~\eqref{57} we use
the standard notations for the basic hypergeometric series given in~\cite{ga}. Then it can be
proved (see \cite{ter2} for details)
that $p_n$ is an eigenstate of the operator $A$
\begin{gather*}
  Ap_n=(q^n +q^{-1}abcdq^{-n})p_n, \qquad n=0,1,2,\dots.
\end{gather*}
The operator $A^*p_n=xp_n$ is tridiagonal and its matrix elements
in the basis
\begin{gather}
 (p_0(x),p_1(x),p_2(x),\dots)
\label{59}
\end{gather}
are obtained from
\begin{gather*}
 xp_n=b_np_{n+1}+a_np_n+c_np_{n-1}, \qquad p_{-1}=0
\end{gather*}
which is the
three term recurrence relation \cite {ask} of the Askey--Wilson polynomials.
Let $\mathcal{A}$ denote the tridiagonal matrix
\begin{gather*}
\mathcal{A} =
\begin{pmatrix} a_0 & c_1         \\
                b_0 & a_1 &  c_2   \\
                    & b_1 &  a_2 &\cdot   \\
                     &    &  \cdot & \cdot
\end{pmatrix}
\end{gather*}
representing the operator $A^*$ in the basis \eqref{59}.
The spectrum of the boundary operators is obtained by the
correspondingly rescaled shifts. We will summarize the results. In the
representation $\pi$ with basis
\begin{gather*}
  (p_0(x),p_1(x),p_2(x),\dots)^t
\end{gather*}
the operator $D_1-\frac {\delta}{\beta} D_0$ is diagonal
with diagonal eigenvalues
\begin{gather}
 \lambda_n= \frac {1}{1-q}\left (bq^{-n} + dq^{n}
  +1+bd \right )
\label{61a}
\end{gather}
and the operator $D_0-\frac {\gamma}{\alpha}D_1$ is tridiagonal
\begin{gather}
\pi \left(D_0-\frac {\gamma}{\alpha}D_1\right) =
   \frac {1}{1-q}\left (b\mathcal{A}^t + 1+ac \right ).
\label{62}
\end{gather}
In the dual representation $\pi^*$ with respect to the basis \eqref{59}
the operator $D_0-\frac {\gamma}{\alpha}D_1$ is diagonal with eigenvalues
\begin{gather}
\lambda^*_n=\frac {1}{1-q}\left (aq^{-n} + cq^{n}
  +1+ac \right )
\label{64}
\end{gather}
and the operator $D_1-\frac {\delta}{\beta} D_0$ is tridiagonal
\begin{gather}
\pi (D_1-\frac {\delta}{\beta} D_0) =
   \frac {1}{1-q}\left (a\mathcal{A} + 1+bd \right )
\label{65}
\end{gather}
The Askey--Wilson parameters $a$, $b$, $c$, $d$ on the RHS of equations \eqref{61a}, \eqref{62} and \eqref{64}, \eqref{65}
depend on the four boundary rates $\alpha$, $\beta$, $\gamma$, $\delta$.

The left and right boundary vectors  can be chosen (up to a normalization) as
\begin{gather*}
 \langle w \vert = (p_0,0,0,\dots), \qquad
 \vert v \rangle = (p_0,0,0,\dots)^t.
\end{gather*}
Then, the solutions to the eigenvalue equations determine uniquely, in
the considered representation of the boundary operators, the relation of
the boundary rates to the four Askey--Wilson parameters. The explicit
expressions read
\begin{gather*}
a=\kappa^{*}_+,\qquad b=\kappa_+, \qquad
c=\kappa^{*}_-, \qquad d=\kappa_-,
\end{gather*}
where
\begin{gather*}
 \kappa_{\pm}= \frac {-(\beta +\delta -(1-q)) \pm
  \sqrt {(\beta +\delta -(1-q))^2 -4\beta \delta}}{2\beta},\nonumber \\
 \kappa^*_{\pm}=\frac {-(\alpha + \gamma -(1-q)) \pm
   \sqrt {(\alpha +\gamma -(1-q))^2 -4\alpha \gamma}}{2\alpha}.
\end{gather*}
We note the dif\/ferent relations compared to the corresponding ones in \cite{wa}.
The boundary symmetry has rich algebraic properties and depending on
its representation one has various identif\/ications of the boundary rates dependence
with the parameters of the Askey--Wilson polynomials.

We thus conclude: The quadratic algebra of the bulk dif\/fusion process
def\/ine a quantum plane with the $SU_q(2)$ action as its symmetry. The
boundary processes amount to the presence of linear terms in the quadratic
algebra and lead to a reduction of the bulk symmetry. The Askey--Wilson
algebra arises as a symmetry of the boundary problem.
The boundary vectors are expressed in terms of the Askey--Wilson polynomials
whose four parameters are related to the four boundary probability rates.

\subsection*{Acknowledgments}

The author would like to thank the organizers for the invitation to
participate the O'Raifeartaigh symposium and for the warm and friendly
atmosphere during the stay in Budapest.

\pdfbookmark[1]{References}{ref}
\LastPageEnding


\begin{thebibliography}{99}

\footnotesize\itemsep=0pt

\bibitem{schu1}Schuetz G.M., Phase transitions and
critical phenomena, Vol.~19, Academic Press, London, 2000.

\bibitem{schre}Schreckenberg M., Schadschneider A.,
Nagel K., Ito N., Discrete stochastic models for traf\/f\/ic f\/low,
{\it Phys. Rev.~E} {\bf 51} (1995), 2939--2949, \href{http://arxiv.org/abs/cond-mat/9412045}{cond-mat/9412045}.

\bibitem{mac} Macdonald J.T., Gibbs J.H.,
Pipkin A.C., Kinetics of biopolymerization on nucleic acid templates,
 {\it Biopolymers} {\bf 6} (1968), 1--25.
 
\bibitem{spo} Krug J., Spohn H., Kinetic roughening of growing surfaces, in Solids
Far from Equilibrium, Editor C.~Godreche, Cambridge University Press, Cambridge, 1991, 412--525.

\bibitem{schu2}Sandow S., Schuetz G.M., On $U_q(SU(2))$-symmetric driven dif\/fusion, {\it Europhys. Lett.}
{\bf 26} (1994), 7--13, \mbox{\href{http://arxiv.org/abs/cond-mat/9307027}{cond-mat/9307027}}.

\bibitem{der1}Derrida B., Evans M.R., Hakim V.,
Pasquier V., Exact solution of a 1D asymmetric exclusion model using a matrix formulation,
{\it J. Phys.~A: Math. Gen.} {\bf 26} (1993), 1493--1517.

\bibitem{der2}Derrida B., An exactly soluble non-equilibrium system: the asymmetric simple exclusion process,
{\it Phys. Rep.} {\bf 301} (1998), 65--83 (and references therein).

\bibitem{do}Schuetz G.M., Domany E., Phase transitions in an exactly soluble one-dimensional asymmetric
exclusion model, {\it J. Stat. Phys.} {\bf 72} (1993), 277--296, \href{http://arxiv.org/abs/cond-mat/9303038}{cond-mat/9303038}.

\bibitem{kru}Krug J., Boundary-induced phase transitions in driven dif\/fusive systems,
{\it Phys. Rev. Lett.} {\bf 67} (1991), 1882--1885.

\bibitem{r}Isaev A., Pyatov P., Rittenberg V., Dif\/fusion algebras, {\it J. Phys.~A: Math. Gen.} {\bf 34} (2001), 5815--5834, \href{http://arxiv.org/abs/cond-mat/0103603}{cond-mat/0103603}.
    
\bibitem{an}Aneva B., The noncommutative space of stochastic dif\/fusion systems,
{\it J.~Phys.~A: Math. Gen.} {\bf 35} (2002), 859--877.

\bibitem{ess}Essler F.H.L., Rittenberg V., Representations of the quadratic algebra and
partially asymmetric dif\/fusion with open boundaries, {\it J.~Phys.~A: Math. Gen.} {\bf 29} (1996), 3375--3407,
\href{http://arxiv.org/abs/cond-mat/9506131}{cond-mat/9506131}.

\bibitem{ar} Arndt P.F., Heinzel T., Rittenberg V., Stochastic models on a ring and
quadratic algebras. The three-species dif\/fusion problem, {\it J.~Phys.~A: Math. Gen.} {\bf 31} (1998), 833--843,
\href{http://arxiv.org/abs/cond-mat/9703182}{cond-mat/9703182}.

\bibitem{za}Fairlie D., Zachos C., Multiparameter associative generalizations of canonical
commutation relations and quantized planes, {\it Phys. Lett.~B} {\bf 256} (1991), 43--49.

\bibitem{we}Wess J., Zumino B., Covariant dif\/ferential calculus on the quantum hyperplane,
{\it Nucl. Phys.~B, Proc. Suppl.} {\bf 18} (1990), 302--312.

\bibitem{ev}Blythe R.A., Evans M.R., Colaiori F.,
Essler F.H.L., Exact solution of a partially asymmetric exclusion model using a deformed
oscillator algebra, {\it J. Phys.~A: Math. Gen.} {\bf 33} (2000), 2313--2332.

\bibitem{san}Sandow S., Partially asymmetric exclusion process with open boundaries,
{\it Phys. Rev. E} {\bf 50} (1994), 2660--2667, \href{http://arxiv.org/abs/cond-mat/9405073}{cond-mat/9405073}.

\bibitem {sti}Stinchcombe R.B., Schuetz G.M., Application of operator algebras to stochastic dynamics
and the Heisenberg chain, {\it Phys. Rev. Lett.} {\bf 75} (1995), 140--143.

\bibitem{aco}Arik M., Coon D.D., Hilbert spaces of analytic functions and generalized coherent states,
{\it J.~Math. Phys.} {\bf 17} (1976), 524--527.

\bibitem{ku}Kulish P.P., Damaskinsky E.V., On the $q$ oscillator and the quantum algebra $su_q(1,1)$,
{\it J.~Phys.~A: Math. Gen.} {\bf 23} (1990), L415--L419.

\bibitem{chai}Chaichian M., Kulish P.P., Quantum Lie superalgebras and $q$-oscillators,
{\it Phys. Lett.~B} {\bf 234} (1990), 72--80.

\bibitem{sa}Sasamoto T., One-dimensional partially asymmetric simple exclusion
process with open boundaries: ortho\-go\-nal polynomials approach,
{\it J.~Phys.~A: Math. Gen.} {\bf 32} (1999), 7109--7131.

\bibitem{wa}Uchiyama M., Sasamoto T., Wadati M., Asymmetric simple exclusion process
with open boundaries and Askey--Wilson polynomials,
{\it J.~Phys.~A: Math. Gen.} {\bf 37} (2004), 4985--5002, \href{http://arxiv.org/abs/cond-mat/0312457}{cond-mat/0312457}.

\bibitem{zhe1}Zhedanov A.S., Hidden symmetry of Askey--Wilson polynomials,
{\it Teoret. Mat. Fiz.} {\bf 89} (1991), 190--204.

\bibitem{zhe2} Granovskii Y.A., Lutsenko I.M., Zhedanov  A.S., Mutual integrability, quadratic algebras
and dynamical symmetry, {\it Ann. Physics} {\bf 217} (1992), 1--20.

\bibitem{ter1}Terwilliger P., An algebraic approach to the Askey scheme of orthogonal polynomials,
{\it Lecture Notes in Mathematics}, Vol.~1883, Editors F.~Marcellan and W.V.~Assche, Springer, Berlin, 2006, 225--330.

\bibitem{ter2}Terwilliger P., Two relations that generalize the $q$-Serre relations and the Dolan--Grady relations,
Proceedings of the Nagoya 1999 International Workshop on Physics and Combinatorics, Editors
A.N. Kirillov, A.~Tsuchiya and H.~Umemura, World Sci. Pub., River Edge, NJ, 2001, 377--398,
\href{http://arxiv.org/abs/math.QA/0307016}{math.QA/0307016}.

\bibitem{dol}Dolan L., Grady M., Conserved charges from self-duality,
{\it Phys. Rev.~D}{\bf 25} (1982), 1587--1604.

\bibitem{ask}Askey R.A., Wilson J.A., Some basic hypergeometric orthogonal polynomials that
generalize Jacobi polynomials,
{\it Mem. Amer. Math. Soc.} (1985), no.~319, 55 pages.

\bibitem{ga}Gasper G., Rahman M., Basic hypergeometric series,  {\it Encyclopedia of Mathematics and Its
Applications}, Vol.~35, Cambridge University Press, Cambridge, 1990.
\end{thebibliography}
\end{document}